\documentclass{aa}
\usepackage{graphicx}
\begin{document}
\title{Force-free pulsar magnetosphere: instability and generation of MHD 
waves} 
\author{V.~Urpin\inst{1,2}}
\institute{$^{1)}$ INAF, Osservatorio Astrofisico di Catania,
           Via S.Sofia 78, 95123 Catania, Italy \\
           $^{2)}$ A.F.Ioffe Institute of Physics and Technology and
           Isaac Newton Institute of Chile, Branch in St. Petersburg,
           194021 St. Petersburg, Russia
}
\date{\today}

\abstract
{Magnetohydrodynamic (MHD) instabilities can play an important role in the 
structure and dynamics of the pulsar magnetosphere.
}
{We consider the instability caused by differential rotation that is
suggested by many theoretical models. 
} 
{Stability is considered by means  of a linear analysis within the frame
of the force-free MHD.
} 
{We argue that differentially rotating magnetospheres are unstable 
for any particular geometry of the magnetic field and rotation law. The
characteristic growth time of instability is of the order of the rotation 
period. The instability can lead to fluctuations of the emission and 
enhancement of diffusion in the magnetosphere.
}
{}


\titlerunning{Instability of the pulsar magnetosphere}

\maketitle

\section{Introduction} 

A global structure of the pulsar magnetosphere is the key question 
to answer for understanding the energy outflow to the exterior. Likely, the 
magnetospheres consist of electron-positron plasma with some 
amounts of ions. This plasma can affect the radiation produced 
in the inner region of the magnetosphere or at the stellar surface. 
Therefore, understanding the properties of a magnetosphere is of 
crucial importance for the interpretation of observations. 
In recent years, some progress has been achieved in theoretical 
models of the pulsar magnetosphere (see, e.g., Goodwin et al. 2004; 
Contopoulos et al. 1999; Komissarov 2006). Apart from a quasi-static 
structure, however, various non-stationary phenomena (such as waves,
instabilities, etc.) can play an important role. They also may affect 
the radiation  but, perhaps, idealised quasi-static magnetospheric 
models cannot be valid in the presence of physical instabilities. 
For example, the electrostatic oscillations with a low frequency have 
been considered recently by Mofiz et al. (2011), who found that the 
thermal and magnetic pressures can generate oscillations that 
propagate near the equator. These low-frequency electromagnetic waves 
are of central importance for understanding the underlying processes 
in the formation of the radio spectrum (see, e.g., Melrose 1996 and 
reference therein).

Instabilities of magnetohydrodynamic modes can also occur in the 
pulsar magnetosphere. For typical values of the magnetic field, the 
electromagnetic energy density is much greater than the kinetic and 
thermal energy density. This suggests that for much of the 
magnetosphere, the force-free condition is a good approximation for 
determining the magnetic field. Magnetohydrodynamic (MHD) processes under 
this condition are poorly studied but they might be very particular. 
One of the MHD phenomena that can occur in the pulsar magnetosphere 
is the so-called diocotron instability, which is the non-neutral 
plasma analog of the Kelvin-Helmholtz instability. This instability 
has been studied extensively in the context of laboratory plasma 
devices (see, e.g., Levy 1965; Davidson 1990; Davidson \& Felice 
1998). The existence around pulsars of a differentially rotating 
equatorial disc with non-vanishing charge density could trigger a 
shearing instability of diocotron modes (Petri et al. 2002). In 
the non-linear regime, the diocotron instability can cause diffusion 
of the charged particles across the magnetic field lines outwards 
(Petri et al. 2003). The role of a diocotron instability in causing 
drifting subpulses in radio pulsar emission has been considered by 
Fung et al. (2006). Note that the diocotron modes should be 
substantially suppressed in a neighbourhood of the light cylinder 
where relativistic effects become important (Petri 2007). 

Recently, one more mode of magnetospheric oscillations has been 
considered by Urpin (2011). This mode is closely related to the
Alfven\'enic waves of the standard magnetohydrodynamics 
modified by the force-free condition and non-vanishing electric 
charge density. Like Alfv\'en waves, the magnetospheric waves of
a small amplitude are transverse (plasma motions are 
perpendicular to the wavevector). The magnetospheric waves 
can be unstable because there is a number of destabilising 
factors in the magnetosphere (differential rotation, electric 
currents, non-zero charge density, etc.). The electric current 
usually provides a destabilising effect that leads to the so-called 
Tayler instability (see, e.g., Tayler 1973a, b). This instability
is well studied in both laboratory and stellar conditions. It arises 
on the Alfv\'en time scale and is particularly efficient if the 
strengths of the toroidal and poloidal field components differ 
substantially (see, e.g., Bonanno \& Urpin 2008a.b). This condition 
can be fulfilled in many magnetospheric models (see, e.g., 
Contopoulos et al. 1999).

Various models of the magnetosphere predict that rotation should be 
differential (e.g., Mestel \& Shibata 1994; Contopoulos et al.1999). 
It is known that differential rotation in combination with the 
magnetic field leads to the magnetorotational instability (Velikhov 
1959; Balbus \& Hawley 1991). This instability may occur even in a 
very strong magnetic field (Urpin \& R\"udiger 2005). However, the 
instability in the pulsar magnetosphere can differ qualitatively from 
the standard magnetorotational instability because of non-vanishing 
charge density and the force-free condition. In this paper, we consider 
the instability in the pulsar magnetosphere caused by differential 
rotation.

\section{Basic equations}

Despite uncertainties in estimates of many parameters, plasma in the 
pulsar magnetosphere is likely collisional and the Coulomb mean free 
path of particles is shorter than the characteristic length scale. 
Therefore, the MHD description can be applied to such plasma. The 
partial MHD momentum equations for the electrons and positrons can 
be obtained by multiplying the Boltzmann kinetic equation by the 
velocity and integrating over it. Assuming that plasma is 
non-relativistic, the momentum equation for particles of the sort 
$\alpha$ ($\alpha = e, p$) reads
\begin{eqnarray}
m_{\alpha} n_{\alpha} \left[
\dot{{\bf V}}_{\alpha}
+ ({\bf V}_{\alpha} \cdot \nabla) {\bf V}_{\alpha} \right]
= - \nabla p_{\alpha} 
+ n_{\alpha} {\bf F_{\alpha}} +
\nonumber \\
e_{\alpha} n_{\alpha} \left({\bf E} + \frac{{\bf V}_{\alpha}}{c}
\times {\bf B} \right) + {\bf R}_{\alpha}
\end{eqnarray}  
(see, e.g., Braginskii 1965, where the general plasma formalism
is developed); the dot denotes the partial time 
derivative. Here, ${\bf V}_{\alpha}$ is the mean velocity of 
particles $\alpha$, $n_{\alpha}$ and $p_{\alpha}$ are their 
number density and pressure, respectively, ${\bf F}_{\alpha}$ 
is an external force acting on the particles $\alpha$ (in our 
case ${\bf F}_{\alpha}$ is the gravitational force), ${\bf E}$ 
and ${\bf B}$ are the electric and magnetic fields, respectively: 
${\bf R}_{\alpha}$ is the internal friction force caused by 
collisions of the particles $\alpha$ 
with other sorts of particles. Since ${\bf R}_{\alpha}$ is the 
internal force, the sum of ${\bf R}_{\alpha}$ over $\alpha$ is 
zero in accordance with Newton's third law. Hence, we have
in the electron-positron plasma ${\bf R}_e = - {\bf R}_p$.

A calculation of ${\bf R}_{\alpha}$ is a very complicated 
problem of plasma physics but we will obtain it using simple
physical arguments. Generally, ${\bf R}_{\alpha}$ is proportional 
to the difference of partial velocities of particles $({\bf V}_e - 
{\bf V}_p)$ and to the temperature gradient (see, e.g., Braginskii 
1965). We will neglect thermal diffusion in ${\rm R}_{\alpha}$ 
because it is usually small in astrophysical conditions and take 
into account only friction caused by a difference in the partial 
velocities. Then,
\begin{equation}
{\bf R}_e = - \frac{m_e n_e}{\tau_e} ({\bf V}_e - {\bf V}_p),
\end{equation}
where $\tau_e$ is the relaxation time of electrons. Note 
that this simple expression for ${\bf R}_e$  is often used 
even in a laboratory plasma (Braginskii 1965) and 
yields qualitatively correct results. We assume that accuracy 
of Eq.(2) is sufficient in the pulsar magnetosphere as well. 

In the electron-positron plasma, both sorts of particles have a 
small mass and the inertial force can be neglected in Eqs.(1). 
The gravitation force is also weak because of a small mass, and 
the gas pressure is much smaller than the magnetic pressure. Hence, 
both these forces can be neglected in Eq.(1) as well. Then, we have
\begin{equation}
e_{\alpha} n_{\alpha} \left({\bf E} + \frac{{\bf V}_{\alpha}}{c}
\times {\bf B} \right) + {\bf R}_{\alpha} = 0.
\end{equation}  
It is more convenient to use linear combinations of
Eq.(3) rather than to solve partial equations. Let us define 
the hydrodynamic velocity and electric current as
\begin{equation}
{\bf V} = \frac{1}{n} (n_e {\bf V}_e + n_p {\bf V}_p) , \;\;\;
{\bf j}= e (n_p {\bf V}_p - n_e {\bf V}_e),
\end{equation}
where $n=n_e + n_p$. Then, the velocities of the electrons and 
positrons are
\begin{equation}
{\bf V}_e = \frac{1}{2 n_e} \left( n {\bf V} - \frac{{\bf j}}{e} \right), 
\;\;\;
{\bf V}_p = \frac{1}{2 n_p} \left( n {\bf V} + \frac{{\bf j}}{e} \right).
\end{equation}  
If $n$ is much greater than the charge number density 
$|n_p - n_e|$, we have $V \gg j/en$. However, this inequality cannot 
be valid if $|n_p - n_e| \sim n$. In the general case, the sum of 
electron and positron Eqs.(3) yields the equation of hydrostatic equilibrium 
\begin{eqnarray}
\rho_e {\bf E} + \frac{1}{c} \; {\bf j} \times {\bf B} = 0,
\end{eqnarray} 
where $\rho_e = e (n_p - n_e) = e \delta n$ is the charge density.
Taking the difference of Eq.(3) for electrons and positrons,
we obtain
\begin{equation}
{\bf j} = \rho_e {\bf V} + \sigma \!\left({\bf E} \! + \!
\frac{{\bf V}}{c} \! \times \! {\bf B} \right), 
\end{equation}
where $\sigma = e^2 n_p \tau_e/m_e$ is the conductivity of 
plasma.

Eqs.(6) and (7) can be written as
\begin{equation}
\rho_e {\bf E}^* + \frac{1}{c} \; {\bf J} \times {\bf B} = 0\; ,
\;\;\;\;\;\;\;
{\bf J} -  \sigma {\bf E}^* =0 , 
\end{equation}
where
\begin{equation}
{\bf E}^* = {\bf E} + 
\frac{{\bf V}}{c} \! \times \! {\bf B} \; , \;\;\;\; {\bf J} =
{\bf j} -\rho_e {\bf V}.
\end{equation}
Eliminating ${\bf E}^*$ from 
Eq.(8) in favour of ${\bf J}$, we have
\begin{equation}
{\bf J} + \frac{\sigma}{c \rho_e} {\bf J} \times {\bf B}
= 0.
\end{equation}
It follows immediately from this equation that 
${\bf J}_{\parallel} = 0$. Calculating the cross production
of Eq.(10) and ${\bf B}$, we obtain ${\bf J}_{\perp} =
0$ as well. Then, second Eq.(8) yields ${\bf E}^* =0$. Hence, 
Eqs.(6)-(7) are equivalent to the conditions 
\begin{equation}
{\bf j} = \rho_e {\bf V}\;, \;\;\;
{\bf E} =  
- \frac{{\bf V}}{c} \times {\bf B}.
\end{equation}
This  implies that the force-free condition in combination 
with Ohm's law (Eq.(7)) is equivalent to the condition of 
the frozen-in magnetic field and the presence of only advective 
currents. Or, in other words, the force-free condition and Ohm's 
law are compatible only if the electric current is poorly advective 
and the magnetic field is frozen-in. Note that this statement 
is valid at any relation between the electron and positron number
densities.

\section{Instability in the pulsar magnetosphere}

The set of MHD equations complemented by the Maxwell equations   
reads in the force-free pulsar magnetosphere
\begin{eqnarray}
\nabla \cdot {\bf E} = 4 \pi \rho_e , \;\;\; \nabla \times
{\bf E} = - \frac{1}{c} \frac{\partial {\bf B}}{\partial t} ,
\nonumber \\
\nabla \cdot {\bf B} = 0 ,\;\;\; 
\nabla \times {\bf B} = \frac{1}{c} 
\frac{\partial {\bf E}}{\partial t} + \frac{4 \pi}{c} {\bf j}, 
\nonumber \\
{\bf j} \approx \rho_e {\bf V}, \;\;\;
{\bf E} \approx  - \frac{{\bf V}}{c} \times {\bf B}.
\end{eqnarray}  
Note one important property of steady state magnetospheres 
($\partial / \partial t = 0$). Such magnetospheres can exist 
only if the hydrodynamic velocity is non-vanishing, ${\bf V} 
\neq 0$. Indeed, let us assume that ${\bf V}=0$. Then, we 
have from Eqs.(12) that ${\bf j}$ and ${\bf E}$ are equal to 
zero. If the electric field is zero then $\rho_e$ 
is also vanishing. Since ${\bf j}=0$ the magnetic field has a 
vacuum structure ($\nabla \cdot {\bf B} = 0$, $\nabla \times 
{\bf B} = 0$), which means that the magnetosphere does not exist at 
all. The conclusion that there should exist hydrodynamic flows 
in the magnetosphere is the intrinsic property of the equations 
of the force-free magnetohydrodynamics and is valid at any 
relation between the electron and positron number densities.  

The MHD processes governed by Eq.(12) are very particular, which 
point can be illustrated by considering a linear instability.
We assune that the electric and magnetic fields are equal to 
${\bf E}_0$ and ${\bf B}_0$ in the unperturbed magnetosphere. 
The unperturbed charge density and velocity are $\rho_{e0}$ and 
${\bf V}_0$, respectively. For the sake of simplicity, we assume 
that motions in the magnetosphere are non-relativistic, $V_0 \ll c$. 
Linearising Eqs.(12), we can obtain the set of equations that 
describes the behaviour of modes with a low amplitude. Small 
perturbations will be indicated by subscript 1. For the sake 
of simplicity, we treat instability of an 
axisymmetric magnetosphere with respect to axisymmetric 
perturbations. We consider perturbations with a short 
wavelength and space-time dependence $\propto \exp(i \omega t 
- i {\bf k} \cdot {\bf r})$ where $\omega$ and ${\bf k}$ are 
the frequency and wave vector, respectively; the wave vector 
${\bf k}$ has no $\varphi$-component. A short wavelength 
approximation applies if the wavelength of perturbations, $\lambda 
= 2 \pi /k$, is short compared to the characteristic length 
scale of the magnetosphere, $L$. 
Note that, generally, the instability criteria for short 
wavelength perturbations can differ from those for global modes 
with the lengthscale comparable to $L$. Instability of 
global modes is usually sensitive to details of the global 
magnetospheric structure and boundary conditions, which are quite 
uncertain in the pulsar magnetosphere. In contrast, the instability 
of short wavelength perturbations is entirely determined by local 
characteristics of the magnetosphere, which are less uncertain. 
Note also that the boundary conditions and instability of global 
modes can seriously modify a non-linear development of short 
wavelength perturbations, particularly if the global modes grow 
faster than the short wavelength modes. However, in this paper we 
consider only a linear stage of instability.

As explained above, instability can occur because of either 
differential rotation or electric currents. The structure of a 
pulsar magnetosphere and its magnetic topology is quite uncertain 
even in the axisymmetric model. Therefore, we consider in this 
paper only instability caused by differential rotation and neglect 
effects associated to electric currents. We will show that 
instability caused by differential rotation can arise at any 
magnetic geometry.

Substituting the frozen-in condition ${\bf E} =  - {\bf V} \times 
{\bf B}/c$ into the equation $c \nabla \times {\bf E} = - \partial 
{\bf B}/ \partial t$ (Eq.(12)) and linearising the obtained
induction equation, we have  
\begin{equation}
i \omega {\bf B}_1 = \nabla \times ( {\bf V}_1 \times {\bf B}_0
+ {\bf V}_0 \times {\bf B}_1 ).
\end{equation}
If the unperturbed velocity is caused mainly by rotation, 
${\bf V}_0 = s \Omega {\bf e}_{\varphi}$ where $s$ is the
cylindrical radius and ${\bf e}_{\varphi}$ the unit vector
in the $\varphi$-direction, then
\begin{equation}
i \omega {\bf B}_1 - s {\bf e}_{\varphi} ({\bf B}_1 \! \cdot 
\! \nabla \Omega) \!=\! 
i {\bf B}_0 ({\bf k} \cdot {\bf V}_1 \!)
- \! i {\bf V}_1 ({\bf k} \cdot {\bf B}_0 \!).
\end{equation}
This is a standard equation that describes perturbations of the
magnetic field in different types of the magnetorotational
instability (see, e.g., Balbus \& Hawley 1991).

Substituting ${\bf j}$ from Eq.(11) into the Maxwell equation in 
the second line of Eq.(12) and linearising it, we obtain 
\begin{equation}
4 \pi \rho_{e0} {\bf V}_1 = - \left[ i (c {\bf k} \times
{\bf B}_1 + \omega {\bf E}_1) + 4 \pi \rho_{e1} {\bf V}_0 \right].
\end{equation}
Using the linearised frozen-in condition and neglecting terms 
$\sim (V_0/c)$, this expression can be transformed into 
\begin{equation}
{\bf V}_1 + \frac{i \omega}{4 \pi c \rho_{e0}} {\bf B}_{0} \times 
{\bf V}_1 = - \frac{i c}{4 \pi \rho_{e0}} {\bf k} \times {\bf B}_1
- \frac{\rho_{e1}}{\rho_{e0}} {\bf V}_0.
\end{equation}
The perturbation of the charge density can be calculated from
the equation $\rho_{e1} = \nabla \cdot {\bf E}_1 /4 \pi$.
We have with the accuracy in terms of the lowest order in $\lambda
/L$
\begin{equation}
\rho_{e1} = \frac{i}{4 \pi c} [ {\bf B}_0 \cdot 
({\bf k} \times {\bf V}_1 ) -
{\bf V}_0 \cdot ({\bf k} \times {\bf B}_1 )].
\end{equation}
Substituting Eq.(17) into Eq.(16) and neglecting terms 
$\sim V^2/c^2$, we obtain the second equation, coupling 
${\bf B}_1$ and ${\bf V}_1$,
\begin{equation}
4 \pi c \rho_{e0} {\bf V}_1 + i \omega {\bf B}_0 \times {\bf V}_1 
+ i {\bf V}_0 [{\bf B}_0 \cdot ( {\bf k} \times {\bf V}_1 )]
\!=\! - i c^2 {\bf k} \times {\bf B}_1.  
\end{equation}
Two equations (14) and (18) describe the coupled evolution of 
small perturbations of the velocity and magnetic field. 

We study only MHD modes with $\omega \ll ck$ because one 
can split electromagnetic and hydromagnetic modes in this case. 
At $\omega \sim ck$, a consideration becomes cumbersome since
electromagnetic and hydromagnetic modes are strongly coupled.
Apart from this, MHD effects operate basically on a timescale 
longer than the inverse light frequency, $(ck)^{-1}$, because
$V \ll c$ in our model. Therefore, one can expect that MHD 
instability should be particularly efficient for modes with a 
relatively low frequency, $\omega \ll ck$. That is why we consider
such modes first.

Estimating $B_1 \sim V_1 (k B / \omega)$ from Eq.(14), we
obtain that the second and third terms on the l.h.s. of 
Eq.(18) are small compared to the term on the r.h.s. by 
a factor $\omega^2 /c^2 k^2$. Neglecting these terms on the
l.h.s., we have 
\begin{equation}
4 \pi \rho_{e0} {\bf V}_1 = 
 - i c {\bf k} \!\times\! {\bf B}_1 .
\end{equation}
Modes turn out to be transverse, ${\bf k} \cdot {\bf V}_1 
\approx 0$. Subsituting Eq.(19) into Eq.(14), we obtain
\begin{equation}
i \omega {\bf B}_1 - s {\bf e}_{\varphi} ( {\bf B}_1 \cdot
\nabla \Omega ) + \frac{c ({\bf k} \cdot {\bf B}_0)}{4 \pi \rho_{e0}}
{\bf k} \times {\bf B}_1 = 0.
\end{equation}
The dispersion equation can be obtained from Eq.(20) in the 
following way. Calculating a scalar product of Eq.(20) and 
$\nabla \Omega$, we obtain the expression for $({\bf B}_1 \cdot 
\nabla \Omega)$ in terms of $[{\bf B}_1 \cdot ({\bf k} \times 
\nabla \Omega)]$. Substituting this expression into Eq.(20), we 
can express after some algebra ${\bf B}_1$ in terms of $[{\bf B}_1 
\cdot ({\bf k} \times \nabla \Omega)]$. Then, a scalar product of 
the obtained equation and $({\bf k} \times \nabla \Omega)$ yields the 
dispersion relation in the form
\begin{equation}
\omega^2 = \frac{c^4 k^2 ({\bf k} \cdot {\bf b})^2}{\Omega_m^2}
- s \frac{c^2 ({\bf k} \cdot {\bf b})}{\Omega_m} [{\bf e}_{\varphi}
\cdot ( {\bf k} \times \nabla \Omega)],
\end{equation}
where $\Omega_m = 4 \pi c \rho_{e0} / B_0$ and ${\bf b} = {\bf B}_0 
/ B_0$.

If rotation is rigid and $\nabla \Omega =0$, we obtain the
dispersion equation of magnetospheric waves considered by 
Urpin (2011). Note that in our case ${\bf k} \cdot {\bf V}_0 = 0$ 
because ${\bf k}$ has no azimuthal component, whereas ${\bf V}_0$ 
corresponds to rotation and has only the azimuthal component. 
If $|{\bf k} \cdot {\bf b}| > k (\Omega_m |s \nabla \Omega| / 
c^2 k^2)$ the dispersion relation for magnetospheric waves reads
\begin{equation}
\omega = \pm c ( {\bf k} \cdot {\bf b}) \frac{ck}{\Omega_m}.
\end{equation}
Since we assume in our consideration that 
the frequency of magnetohydrodynamic modes should be 
lower than $ck$, the magnetospheric modes exists if 
\begin{equation}
\Omega_m > c ({\bf k} \! \cdot \! {\bf b}).
\end{equation}
This condition can be satisfied for waves with
the wave vector almost (but not exactly) perpendicular to
the magnetic field. If the vector ${\bf k}$ is almost 
perpendicular to ${\bf b}$, it is convenient to denote the 
angle between ${\bf k}$ and ${\bf b}$ in a meridional plane 
as $(\pi/2 -\delta )$. Then, ${\bf k} \cdot {\bf b} = k 
\cos( \pi/2 - \delta ) \approx k \delta \psi$. Then, Eq.(23) 
is satisfied if $\delta < \Omega_m /c k$.

If rotation is differential and $s |\nabla \Omega| > (c^2 k/ 
\Omega_m) ({\bf k} \cdot {\bf b})$, the properties
of magnetospheric waves can be quite different. The first 
term on the r.h.s. of Eq.(21) is always positive and cannot 
lead to instability, but the second term can be negative for
some ${\bf k}$. The instability ($\omega^2 < 0$) is possible 
only if the wavevector is almost perpendicular (but not 
exactly) to the magnetic field and the scalar product 
$({\bf k} \cdot {\bf b})$ is small but non-vanishing. Only 
in this case, the second term on the r.h.s. of Eq.(21) can 
be greater than the first one. Let us estimate the range of 
wave vectors that corresponds to unstable perturbations,
introducing again the angle between ${\bf k}$ and ${\bf b}$ 
as $(\pi/2  - \delta )$. Substituting this 
expression into Eq.(21) and estimating $[{\bf e}_{\varphi} 
\cdot ({\bf k} \times \nabla \Omega)] \sim k |\nabla 
\Omega|$, we obtain that the second term on the r.h.s. of 
Eq.(21) is greater than the first one if
\begin{equation}
\delta  < \frac{s |\Omega_m \nabla \Omega|}{c^2 k^2}.
\end{equation}
The angle $\delta $ turns out to be small, and only 
perturbations with a wave vector almost perpendicular
to ${\bf B}$ can be unstable. 

The instability arises if the second term on the r.h.s. of
Eq.(21) is positive. Therefore, the necessary condition of 
instability reads  
\begin{equation}
\frac{({\bf k} \cdot {\bf b})}{\Omega_m}
[{\bf e}_{\varphi} \cdot ( {\bf k} \times \nabla \Omega)]
> 0.
\end{equation}
Since the sign of $\Omega_m$ depends on the charge 
density, the necessary condition is
\begin{equation}
({\bf k} \cdot {\bf b})
[{\bf e}_{\varphi} \cdot ( {\bf k} \times \nabla \Omega)]
> 0  {\rm or} < 0
\end{equation} 
in the region of positive or negative charge density,
respectively. It turns out that the necessary conditions 
(26) can be satisfied by the corresponding choice of the 
wave vector at any $\nabla \Omega$ and ${\bf b}$. Indeed, 
since ${\bf k}$ is almost perpendicular to the magnetic 
field we can represent it as 
\begin{equation}
{\bf k} \approx \pm k {\bf e}_{\varphi} 
\times {\bf b} + \delta {\bf k},
\end{equation} 
where $\delta {\bf k}$ is a small component of ${\bf k}$ 
parallel (or antiparallel) to ${\bf b}$, $k \gg \delta k$ 
(we neglect terms of the order $(\delta k/k)^2$).
Substituting expression (27) into Eq.(26), we obtain for 
the upper sign with the accuracy in terms linear in 
$\delta k$
\begin{equation}
\pm k (\delta {\bf k} \cdot {\bf b}) ({\bf b} \cdot \nabla \Omega) 
< 0.
\end{equation} 
Obviously, at any sign of $({\bf b} \cdot \nabla 
\Omega)$, one can choose $\delta {\bf k}$ in such a way 
that condition (28) will be satisfied. Condition (26) for 
the region with a negative charge density can be considered 
by analogy. Hence, the necessary condition of instability 
(25) can be satisfied an any differential rotation in the 
regions of positive and negative charge density. 

The characteristic growth rate can be obtained from
Eq.(21), using estimate $({\bf k} \cdot {\bf b}) \sim k
\delta \psi$. Then,
\begin{equation}
|\omega| = \frac{1}{\tau_{\Omega}} \sim |s \nabla \Omega|,
\end{equation}
where $\tau_{\Omega}$ is the growth time of instability
caused by differential rotation. If differential rotation 
is sufficiently strong and $|s \nabla \Omega| \sim \Omega$,
then the growth time of instability is of the order of
the rotation period.

\section{Discussion}

We have considered the instability of a pulsar magnetosphere
caused by differential rotation. The consideration was made
using the force-free approximation that should
be satisfied in the magnetosphere with a high accuracy because 
the electromagnetic energy density is much greater than the 
kinetic and thermal energy density. The main result of this
study is that the differentially rotating force-free magnetosphere
is always unstable. This conclusion is valid for any particular 
magnetic topology and rotation law. The instability considered
in this paper is the representative of a wide class of the 
magnetorotational instabilities (see, e.g., Velikhov 1959; Balbus
\& Hawley 1991; Urpin \& R\"udiger 2005) modified by the force-free
condition and non-vanishing charge density. The typical growth time
of the instability can be quite short and can reach the rotation 
period in the case of a strong differential rotation with $|s \nabla
\Omega| \sim \Omega$. 

Differential rotation is typical for many models of the magnetosphere.
For instance, in the axisymmetric model by Countopoulos et al. (1999)
the angular velocity decreases inversely proportional to the cylindrical
radius beyond the light cylinder and even stronger in front of it.
For this rotation, the growth time of instability should be of the 
order of the rotation period. Numerical simulations by Komissarov 
(2006) showed that within the light cylinder, plasma rotates differentially
basically near the equator and poles. Therefore, a strong differential 
rotation should lead to instability arising in these regions. 
However, the situation can be quite different near the light cylinder
where the instability can occur in a much wider region.

The instability considered can be responsible for fluctuations
of the magnetospheric emission with the characteristic timescale
$\sim 1/\omega$. Hydrodynamic motions accompanying the instability
can be the reason of turbulent diffusion in the magnetosphere.
Note that an influence of the diffusion coefficients should be
strongly anisotropic with a much higher enhancement in the direction
of the magnetic field since the velocity of motions across the 
field is much slower than along it.

Despite the force-free condition that substantially reduces the number 
of modes that can exist in the magnetosphere, there are still many 
destabilising factors that can lead to instability. Apart from
differential rotation, the electric current is likely one more
important factor of destabilisation. Note that the topology of the 
magnetic field can be fairly complicated in the magnetosphere, 
particularly in a region close to the neutron star. This may happen
because the field geometry at the neutron star surface should be
very complex (see, e.g., Bonanno et al. 2005, 2006; Urpin \& Gil 
2004). Therefore, magnetospheric magnetic configurations can be
subject to the so-called Tayler instability caused by a distribution
of currents. This instability in the pulsar magnetosphere will be
considered elsewhere.

\section*{Acknowledgement}
 
The author thanks the Russian Academy of Science for financial
support under the Programme OFN-15.

\end{document}